\documentclass{article}
\usepackage{a4wide}
\usepackage{amsmath}
\usepackage{amsfonts}
\usepackage{amssymb}
\usepackage{psfig}
\usepackage{graphicx}

\newcommand{\rv}{{\bf r}}

\newcommand{\kv}{{\bf k}}
\newcommand{\lv}{{\bf l}}
\title{Optimized Periodic Coulomb Potential in Two Dimension}
\author{Markus Holzmann and Bernard Bernu\\
Laboratoire de Physique Th\'eorique des Liquides\\ UMR
7600 of CNRS\\ Universit\'e Pierre et Marie Curie\\ boite 121, 4 Place
Jussieu\\ F-75252 Paris, France}
\begin{document}
\maketitle
\begin{abstract}
The $1/r$ Coulomb potential is calculated for a two dimensional system with periodic boundary conditions.
Using polynomial splines in real space and a 
summation in reciprocal
space
we obtain
numerically optimized potentials which
allow us efficient calculations
of any periodic (long-ranged) potential up to high precision.
We discuss the parameter space of the optimized potential for the periodic 
Coulomb potential. Compared to the analytic Ewald potential,  
the optimized
potentials can reach higher precisions by
up to several orders of magnitude. We explicitly give
simple expressions for
fast calculations of the periodic Coulomb potential 
where the summation in reciprocal space is reduced to a few terms.
\end{abstract}

\section{Introduction}
Most of classical or quantum simulations
are using periodic boundary conditions to extrapolate the results to the thermodynamic
limit of the bulk.
Typically, boundary conditions
are implemented by including replicas of the original system; the potentials
are then calculated by considering the additional interactions between the particles
in the box with all the periodic images of the replicas. Whereas for short-range potentials the
nearest image convention can often be applied, long-range potentials require an additional
summation over the Fourier components
in reciprocal space 
due to the slow convergence of the contributions of images in real space
(see \cite{revue} for a recent review of how to compute long range potentials within periodic boundary
conditions).

Following Ref.~\cite{Natoli}, we introduce an optimized periodic potential, represented
by two summations, one over Fourier and one over real-space components, which can be obtained
numerically for any potential. We extend the analysis of Ref.~\cite{Natoli} to treat
two-dimensional systems and concentrate particularly on the two-dimensional 
periodic Coulomb potential.

For an arbitrary potential $v(\rv)$, the periodic image potential $v_{pp}(\rv)$ 
is defined by summing over the interactions between one particle in a box
of dimensions $L_x$, $L_y$ and the replicas of the other particles in periodic space,
\begin{equation}
v_{pp}(\rv)=\sum_\lv v(\rv + \lv)=  \sum_\kv \tilde{v}_\kv e^{i \kv \cdot \rv}
\end{equation}
Here, $\lv$ are the Bravais lattice vectors $(n_x L_x, n_y L_y)$ with
$n_x,n_y$ integers, $\tilde{v}_\kv$ are the Fourier components of the potential
summed over all  reciprocal lattice vectors $\kv=2\pi (n_x/L_x,n_y/L_y)$ of
the periodic system.

For a charge $q$, the Coulomb potential is given by
\begin{equation}
v(r)=\frac{q}{r}-\int_{V} d \rv' \frac{q}{|\rv - \rv'|},
\label{Coulomb-rs}
\end{equation}
with Fourier components
\begin{eqnarray}
  \tilde{v}_k =
 \left\{
 \begin{array} {l@{\quad : \quad}l}
 \frac{1}{V} \frac{2 \pi q}{k}  &  k \ne 0 \\
  0 & k=0
\end{array}
\right.
\label{Coulomb-fs}
\end{eqnarray}
in two dimensions where $V=L_x L_y$ is the volume of the box.
A uniform background of opposite charge is subtracted
to enforce charge-neutrality. Since both summations in real space and in reciprocal
space of the periodic potential converge slowly, the standard method is to 
split the periodic potentials into two summations,
\begin{equation}
v_{op}( \rv )= \sum_\lv w( \rv + \lv ) + \sum_{|\kv|\leq K_{c}}  \tilde{y}_\kv e^{i \kv \cdot \rv}
\label{vopt}
\end{equation}
with
\begin{equation}
w(r) \equiv 0 \quad \mathrm{for} \quad r> R_c.
\end{equation}
By definition both summations are converged.
In Ref.~\cite{Natoli}  it has been proposed to use a set of basis functions for
$w(r)$ and determine numerically their coefficients together with $\tilde{y}_\kv$ such
that the difference between the optimized periodic potential $v_{op}( \rv)$ and
the periodic image potential $v_{pp}( \rv )$ is minimized.

An analytical form for the Coulomb potential 
was provided long time ago by Ewald using a Gaussian charge
distribution \cite{ewald}. It gives
\begin{eqnarray}
w^\alpha( r)&=&q\frac{\mathrm{erfc}( \alpha r)}{r}\\
  \tilde{y}_k^\alpha &=&q
 \left\{
 \begin{array} {l@{\quad : \quad}l}
  \frac{2 \pi}{V} \frac{\mathrm{erfc}(k/2 \alpha)}{k} &  k \ne 0 \\
  -\frac{2 \sqrt{\pi}}{\alpha V} & k=0 
\end{array}
\right.
\label{ewald}
\end{eqnarray}
with $\lim_{r\to0}w^\alpha(r)-q/r=-q 2 \alpha/\sqrt{\pi} $
and $\alpha$ is an open parameter which determines the speed of convergence
in both summations.
Due to the exponential convergence of both summations, they
can be truncated, and $R_c$ and $K_c$ can be determined to ensure
any desired precision. Choosing $\alpha=\sqrt{\pi/V}$,
both summations roughly converge equally fast, and $(R_c K_c)$ is the only parameter determining
the precision of the truncated Ewald potential. 
In practice, one typically restricts $R_c < L/2$ with $L=\min \{L_x,L_y\}$ in order to apply the nearest
image convention in real space;
the precision of the potential then relies on $\{\alpha,K_c\}$. 

In Ref.~\cite{Natoli} it has been shown that a numerical fit of the 3D Coulomb potential
reduces considerably the number of terms in k-space  with respect to the analytical Ewald summation
in order to obtain a comparable precisions. This leads
to an important speedup of simulations of charged systems. Further, this method is
not limited to the Coulomb potential, but can easily be applied to any functional form.
This is important in ground state quantum Monte Carlo calculations,
since analytical forms for the potentials  have been shown to provide an accurate description
of the ground state wavefunction of electronic systems \cite{dmc78,backflow};
however, they typically involve
more complicated (long ranged) functions where no easy analytical break-up can be done. 
The optimized potential of Ref.~\cite{Natoli} has the big advantage to be applicable to
all type of function; here, we extend this method to include two-dimensional systems.

In the following section, we shortly remind the basic steps necessary to derive the equations
of the optimized potential which have to be solved numerically, and  give the explicit formulas
for the 2D case. 
The precise numerical evaluation of  Bessel functions and their
integrals are discussed.
In section III are presented the results of the optimized potential.
Explicit simple formulas are given for the 2D periodic Coulomb potential up to a few percents.

\section{Method and formulas for 2D}
\subsection{General method}
The optimized potential, $v_{opt}$,  is determined by minimizing the absolute error 
with respect to the true periodic potential, $v_{PP}$,
\begin{equation}
\chi^{2}=\frac{1}{L^{2}} \int_{L^2} d\rv \left[ v_{pp}(\rv)-v_{op}(\rv) \right]^{2}.
\label{chi2_org}
\end{equation}
Denoting $\tilde{w}_\kv$ the Fourier transform of $w(\rv)$ in the optimized potential,
Eq.(\ref{vopt}) reads
\begin{equation}
v_{op}(\rv) = \sum_\kv e^{i \kv \cdot \rv} \tilde{w}_\kv 
+\sum_{|\kv| \leq K_c} e^{i \kv \cdot \rv} \tilde{y}_\kv,
\end{equation}
and Eq.(\ref{chi2_org}) is split in two sums:
\begin{equation}
\chi^2  =
\sum_{|\kv|\leq K_c} \left( \tilde{v}_\kv-\tilde{y}_\kv- \tilde{w}_\kv \right)^{2}
      +\sum_{|\kv|>K_c} \left( \tilde{v}_{\kv}-\ w_\kv \right)^{2}.
\label{chi2}
\end{equation}
The first term on the rhs of this equation
can be exactly set to zero determining $\tilde{y}_\kv$, 
\begin{equation}
\tilde{y}_\kv= \tilde{v}_\kv - \tilde{w}_\kv \qquad \mathrm{for}\quad |\kv| \leq K_c.
\label{yk1}
\end{equation}
Expanding $w(\rv)=\sum_i t_i c_i(\rv)$  using a set of basis functions $c_i(\rv)$ with Fourier components 
$\tilde{c}_{i\kv}$, Eq.(\ref{yk1}) relates the optimal Fourier coefficients $y_\kv$ to
the optimal coefficients $t_i$ in real space 
\begin{equation}
\tilde{y}_\kv= \tilde{v}_\kv 
- \sum_{i}t_i \tilde{c}_{i\kv} \qquad \mathrm{for}\quad |\kv| \leq K_c.
\label{yk}
\end{equation}
The optimal coefficients $t_i$ can be determined by
minimizing the second term of the rhs of Eq.(\ref{chi2}) leading to  
the following linear equations
\begin{equation}
\sum_n \, \sum_{ |\kv| > K_c} \tilde{c}_{i\kv} \tilde{c}_{n\kv} \, t_n
= \sum_{ |\kv| > K_c} \tilde{v}_\kv \tilde{c}_{i\kv}
\label{LE}
\end{equation}
%
Solving Eq.(\ref{yk}) and Eq.(\ref{LE}) uniquely determines the optimized potential
for any given $R_c$ and $K_c$.

\subsection{Polynomial Basis set}
Here, we use polynomial splines  sitting on a linear grid
with $m$ continuous derivatives as basis set (assuming a spherical symmetry of the potential).
Following Ref.\cite{Natoli},
the splines are defined on intervals $(r_i,r_{i+1})$ with $N_{spline}+1$
equally spaced knots starting at
the origin and ending at $R_c$, $r_i=i \Delta$ with interval $\Delta=R_c/N_{spline}$.
The basis functions are $c_{i\alpha}(r)$ with $0\leq \alpha \leq m$ are defined by imposing
\begin{equation}
\left.\frac{d^\beta c_{i\alpha}(r)}{dr^\beta}\right|_{r=r_{j}}=\delta_{\alpha\beta}\delta_{ij}.
\end{equation}
The divergence of the potential at the origin is explicitly taken in to acount as follow,
\begin{equation}
w(r)=\sum_{i=0}^{N_{spline}} \sum_{\alpha=0}^{m} t_{i\alpha} \frac{c_{i\alpha}(r)}{r^C}
\label{wr}
\end{equation}
where $C=1$  for the Coulomb potential and 
$C=0$ for regular potential at the origin. 
Thus, the basis functions $c_{i\alpha}(r)$ are piecewise polynomial of order $2m+1$
\begin{eqnarray}
  c_{i\alpha}(r) = 
\left\{
 \begin{array} {l@{\quad : \quad}l}
 \Delta^\alpha \sum_{n=0}^{2m+1} S_{\alpha n} \left( \frac{r-r_i}{\Delta} \right)^n  &  r_i<r\leq r_{i+1} \\
 \left(-\Delta\right)^\alpha \sum_{n=0}^{2m+1} S_{\alpha n} \left( \frac{r_i-r}{\Delta} \right)^n  &  r_{i-1}<r\leq r_i 
\end{array}
\right. 
\label{mspline}
\end{eqnarray}
and zero for $|r-r_{i}|>\Delta$.
The constraints at $r=r_i$ fixe half of the $S$-elements
\begin{equation}
S_{\alpha n}=\frac{1}{n!} \delta_{\alpha,n}, \quad \mathrm{for} \quad 0 \leq \alpha,n \leq m.
\end{equation}
The constraints at $r=r_{i\pm1}$ gives
\begin{equation}
  S_{\alpha, n+m+1} =-\sum_{k=0}^\alpha \left(M^{-1}\right)_{kn} \frac{1}{(\alpha-k)!}
\quad \mathrm{for} \quad 0 \leq \alpha, n \leq m
\end{equation}
where $M^{-1}$ is the inverse of the  quadratic matrix 
\begin{equation}
M_{ak}=\frac{(m+1+a)!}{(m+1+a-k)!} \quad \mathrm{for} \quad 0\leq a,k \leq m.
\end{equation}
The required Fourier coefficients $\tilde{c}_{i\alpha k}$ are given by
\begin{equation}
\tilde{c}_{i \alpha k}=\Delta^\alpha \sum_{n=0}^{2m+1}
S_{\alpha n} \left( D^+_{i kn}+(-1)^{\alpha+n} D^-_{i kn} \right)
\label{ciak}
\end{equation}
where
\begin{eqnarray}
D^\pm_{i k n} & = & \pm \frac{1}{V} \int_{r_i}^{r_{i\pm 1}}
d\rv e^{-i \kv \cdot \rv} r^{-C} \left( \frac{r-r_i}{\Delta} \right)^n \\
& = & \pm
\frac{1}{V \Delta^n} \sum_{j=0}^{n} \left({ }^n_j \right) (-r_i)^{n-j} \int_{r_i}^{r_{i\pm 1}}
d\rv \, r^{j-C} e^{-i \kv \cdot \rv}  \\
& = &  \pm
\frac{2 \pi }{V \Delta^n} \sum_{j=0}^{n} \left( { }^n_j \right) (-r_i)^{n-j} \int_{r_i}^{r_{i\pm 1}}
dr \, r^{j+1-C}  J_0(kr)
\label{diak}
\end{eqnarray}
with $J_0(x)$ is the Bessel function of zero order and $\left(Ê{ }^n_j \right)$ denotes the binomial coefficients. The moments of $J_{0}$ can be obtained from its first two moments and a reccurence relation:
\begin{eqnarray}
\int dx J_0(x) & = & x J_0(x)+ \frac{\pi}{2} \left(H_0(x) J_1(x)-H_1(x)J_0(x)\right)\\
\int dx x \, J_0(x) & = & x J_1(x)\\
\int dx x^n J_0(x)&=&x^n J_1(x)+(n-1) x^{n-1} J_0(x) -(n-1)^2 \int dx x^{n-2} J_0(x).
\end{eqnarray}
Here, $J_1$ is the Bessel function of first order and $H_1$ ($H_2$) is Struve's function of
first (second) order.
However, 
contrary to the 3D case, we have not found any ``machine precision'' routine to compute the Struve's function or the integral of $J_{0}$
due to the oscillatory behavior of the integrand.
In Appendix A, we briefly 
describe how to evaluate ``precisely''  the integral of $J_{0}$.

\section{Results for the 2D periodic Coulomb potential}

In this section, we present the results of the optimized potential for the
Coulomb $1/r$ potential in two dimensions, Eq.(\ref{Coulomb-rs}) and Eq.(\ref{Coulomb-fs})
in a square box of length $L=L_x=L_y$.
Hermite splines of fifth order represent the
real space part of the optimized potential insuring two continuous
derivatives ($m=2$ in Eq.\ref{mspline}). 
The $1/r$ divergence at the origin is accounted for by setting $C=1$
with the constraint $t_{i=0,\alpha=0}=q$; symmetry further imposes the
absence of any linear term linear: $t_{i=0,\alpha=2}=0$.
Both constraints can be easily included,
by solving the linear equations
\begin{equation}
\sum_{(j,\beta) \ne \{(0,0),(0,2)\}} A_{i\alpha,j\beta}  t_{j \beta} = b_{i\alpha},
\quad 0\leq \alpha \leq m_{nderv}, \; 0\leq i \leq N_{spline} 
\end{equation}
for 
$(i,\alpha) \ne \{(0,0),(0,2)\}$
with
\begin{equation}
A_{i\alpha,j\beta}=\sum_{k=K_c}^{K_m} \tilde{c}_{i\alpha k} \tilde{c}_{j\beta k}
\quad b_{i \alpha}=\sum_{k=K_c}^{K_m} \left( \tilde{v}_{k} 
- \tilde{c}_{00k} t_{00} \right) \tilde{c}_{i \alpha k}
\label{LE_constr}
\end{equation}
instead of Eq.(\ref{LE}).
Appendix B describes how to impose the Madelung constant of the lattice through the constant term $t_{01}$.

The number of spline intervals, $N_{spline}$, and the cutt-off $K_{c}$ in the reciprocal space
are the open parameters of the model.
In Eq.(\ref{LE_constr}), $K_m$ has to be large enough to insure that each sum is converged.
Since it is an important parameter which determines
the conditioning of the linear equation we first discuss its influence on the stability of
the solution.

The results of the optimized potential are compared with the ``exact'' periodic
Coulomb potential obtained from the Ewald formula including a summation over
many images in real space. Thus machine precision is easily reached for this reference potential. 
Values of the potential are given in units of $q/L$, they are independent
of the size of the box. 

\subsection{Extrapolation of $K_m$}
Results are very sensitive to $K_{m}$ which determines 
the convergence of the matrix elements of $A$ and $b$, Eq.(\ref{LE_constr}). 
From Eq.(\ref{ciak}) and Eq.(\ref{diak}), one finds the dominant
behavior in the limit of $k \to \infty$: $t_{00}\tilde{c}_{00k} - \tilde v_{k} \sim 
  {\cal O}(k^{-7/2})$ and
$\tilde{c}_{i\alpha k} \sim {\cal O}(k^{-7/2})$ for $(i,\alpha) \ne (0,0)$.
Thus the matrix elements in $A$ and $b$,  Eq.(\ref{LE_constr}),  are of
order $k^{-7}$. 
In the large $k$ limit, the truncation therefore introduces an error
of order $K_m^{-5}$.

Considering the asymptotic expansions of $c_{i\alpha k}$ for large $k$, one might 
be able to extend the summation analytically to infinity using
the leading order terms. However, 
in 2D, the difference between the discrete summation over reciprocal lattice vectors
and the continuous integration is comparable to the correction of the continuous integral.
Therefore, we find no improvement by adding those corrections.
Thus, contrary to the 3D case, no analytical continuations are used here.

Next, we consider the stability of the solution with respect to variations of
$K_m$.
Since the short wavelength cut-off destroys the information on small distance behavior of the
potential, we expect that there is a maximum
number of splines $N^{max}_{spline}$, after  which the solution of the linear equation
will become unstable. Roughly, the  resolution in real space is limited
by $\Delta K_m/(2 m+2) >\sim 2 \pi$,
and we get a maximum number of spline knots $N^{max}_{spline}+1$ with
\begin{equation}
N^{max}_{spline} \leq \frac{K_m R_c}{4 \pi (m+1)} 
\label{uncert}
\end{equation}

Using less terms in the summation in reciprocal space, the matrix $A_{i\alpha j\beta}$ in
Eq.(\ref{LE_constr}) becomes ill conditioned. The conditioning of the
linear system can be estimated by comparing the norm 
of the obtained solution $t_i$, $|| {\bf t} ||_\infty=\max\{ |t_i|,i\}$
with $|| {\bf s} ||_\infty=\max\{ |\sum_j A_{ij} t_j-b_i|,i\}$.
If 
$|| {\bf s} ||_\infty/||{\bf t} ||_\infty$ is of order one, the
system is dominated by numerical round-off errors.
This indicates that either the value of
$K_m$ is too small or the number of splines is too large so that no improvement
can be reached by increasing $N_{spline}$ further.

\subsection{Accuracy of the optimized potential}

\begin{figure}
\begin{center}
\resizebox{8.1cm}{!}{\includegraphics{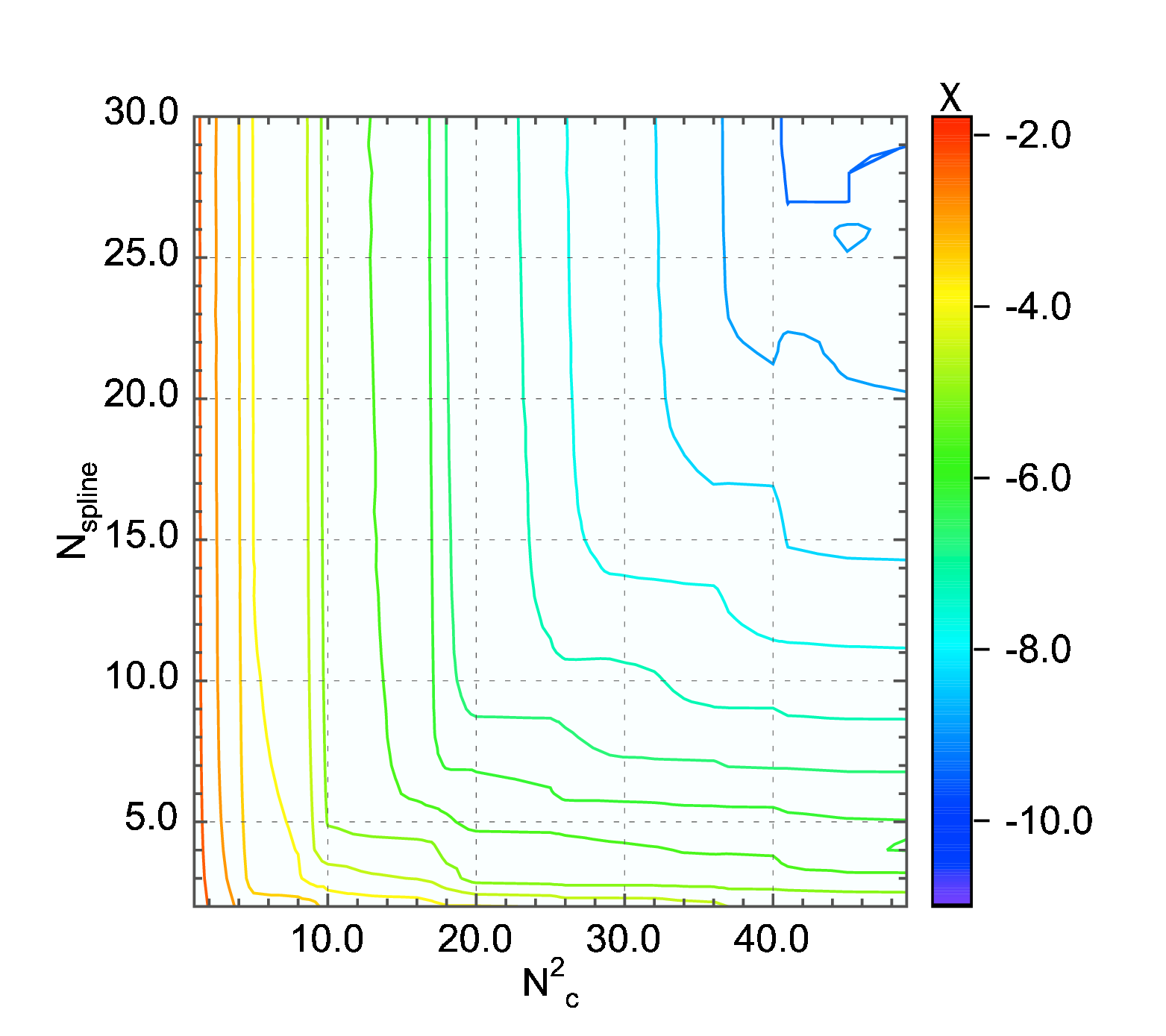}}
\caption[99]{
\label{FIG-2}
Contour plot of the mean error of the  optimized potential versus the number of shells $N^2_{c}$ ($N_{c}=K_{c}L/2\pi$) and the number of splines $N_{splines}$ (see definition of Eq.(\ref{vopt})).
The maximum distance in r-space is chosen to satisfy the
nearest image convention in simulations, $R_c/L=0.5$.
}
\end{center}
\end{figure}

We now study the accuracy of the optimized potential in the $N_{spline}$ - $K_c^2$ -plane
for fixed $R_c/L=0.5$ (nearest image convention). The $\sqrt{\chi^2}$ in units of
$q/L$
corresponds to the average error and is shown in Fig.~\ref{FIG-2}. 
We see that there is an optimum line
in the range of parameters considered. We also note that there
is a difficulty to decrease the precision below $10^{-10}$, mainly because the 
linear system, Eq.(\ref{LE_constr}) becomes more and more ill-conditioned 
as $N_{spline}$ and $K_c^2$ increase.

\begin{figure}
\begin{center}
\resizebox{8.1cm}{!}{\includegraphics{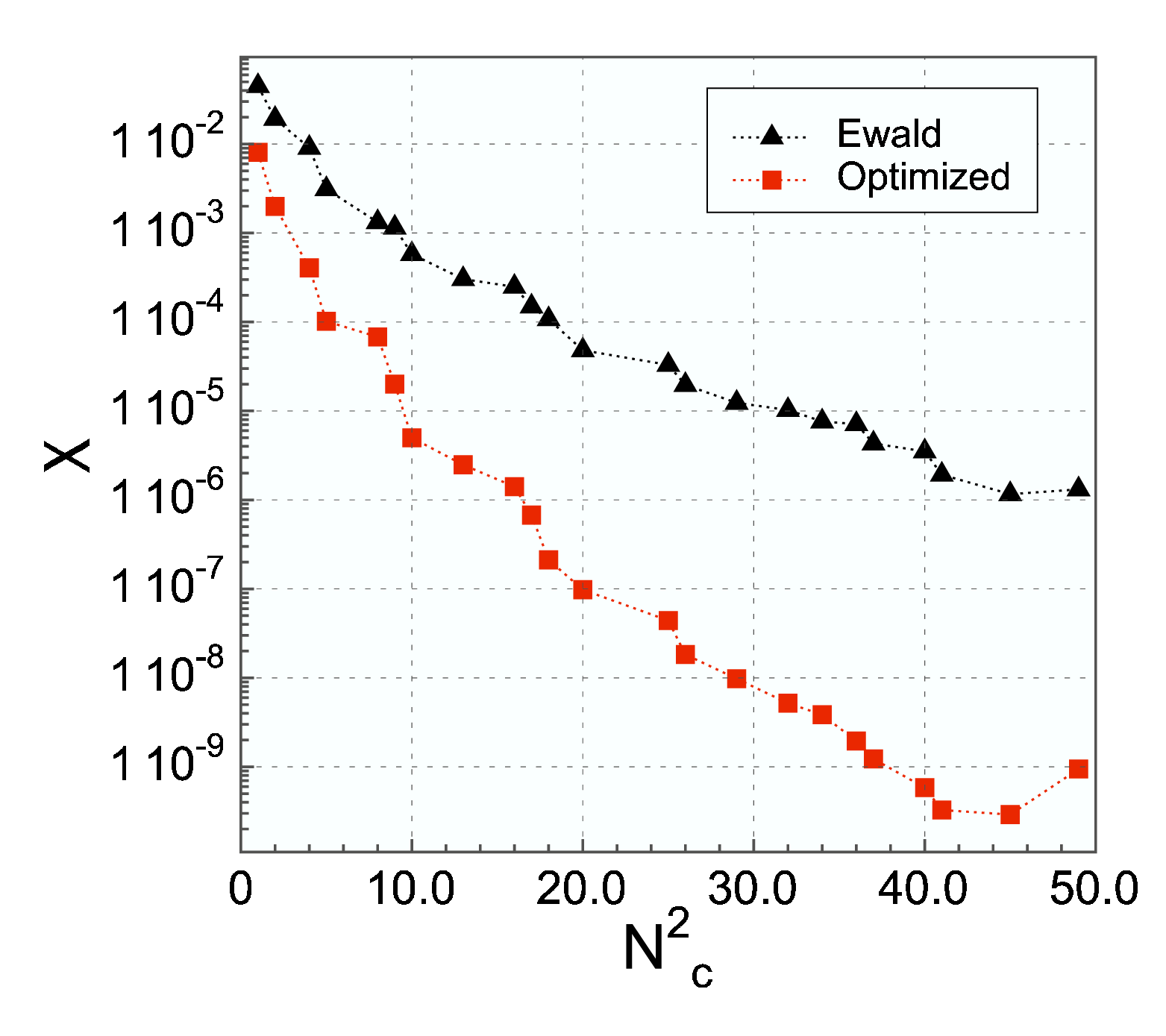}}
\caption[99]{
\label{FIG-1}
Comparison of the precision between the standard Ewald method (with $\alpha=K_c/L$) (triangle) 
and the the optimized potential 
versus $N_{c}=K_{c}L/2\pi$ (square).
The average error $\chi$ is given in units of $q/L$. 
The maximum distance in r-space is chosen to satisfy the
nearest image convention, $R_c/L=0.5$.
}
\end{center}
\end{figure}

Figure~\ref{FIG-1} shows the difference to the true periodic potential for
$N_{spline}=30$ varying $K_c^2$ and compares it with the values of the best nearest image potential
using the analytic Ewald expressions, Eq.(\ref{ewald}) with $\alpha=K_c/L$. For the range of
interest in order to speed up simulations, the optimized potential is always better by at least one order
of magnitude.
However, if very high precision is needed, better than $10^{-10}$, the Ewald method is preferable
to the present optimized procedure, even if it is much more cpu-time consuming, since it has no
stability problems.

The number of splines increases the precision of the potential inside the circle of radius $R_{c}$. Around the corner of the box, only the number of $k$-shells can improve the optimized potential. At fixed number of shells, one reaches rapidly an optimum number of splines after which increasing the number of splines has no more effect. This is seen in Fig.\ref{FIG-1} by the straight vertical lines.
At fixed number of splines, increasing the number of $k$-shells first improves
strongly the optimized potential, but later the improvement almost saturates.
Thus the optimum choice is to take the parameters roughly along the diagonal in Fig.\ref{FIG-1}.

Note that an intermediate precision of $10^{-3}$ to $10^{-6}$ 
is reached with very small values of the optimized potential parameters.
Therefore simple analytical expressions allow us fast evaluations of the periodic Coulomb potential with 
intermediate precisions.

\subsection{Simple expressions for intermediate precision}

\begin{figure}
\begin{center}
\resizebox{8.1cm}{!}{\includegraphics{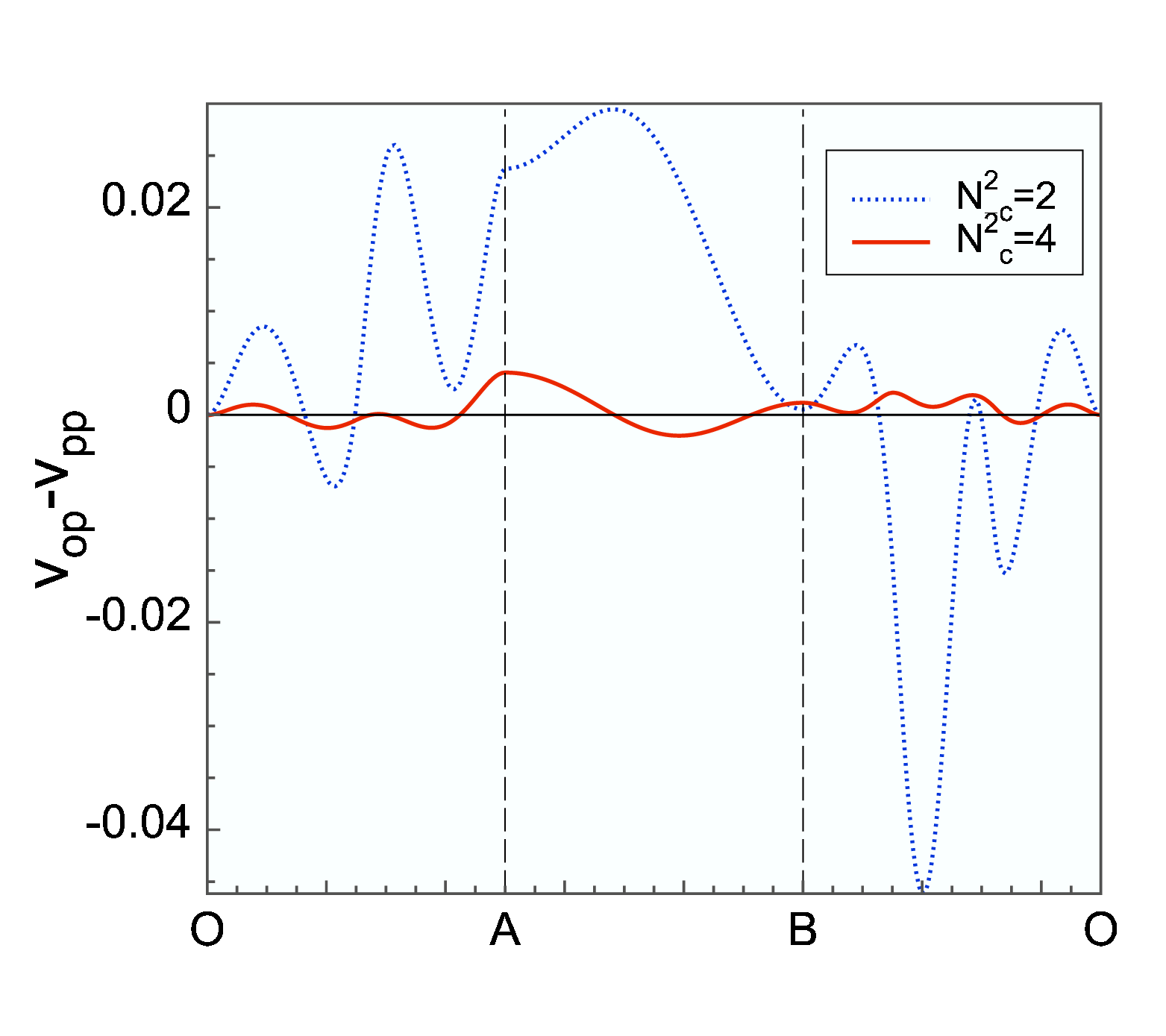}}
\caption[99]{
\label{FIG-3}
Difference between the simple expression of the optimized potential, $v_{op}$,
with the exact Coulomb potential, $v_{pp}$; dooted lined stands for $v_{op}$ using $K_{c}=4\pi/L$, full line for $K_{c}=8\pi/L$.
$O$ stands for the origin, $A$ for the middle of the square side and $B$ for the corner of the square.
}
\end{center}
\end{figure}

If we include only the first five  wavevectors with
$|\kv| \leq K_c=2 \pi/L$ and use $N_{spline}=2$, the minimum to obtain a
smooth curve going to zero at half of the box size, a mean precision of $1\%$ is obtained.
and a maximum error of around $2\%$ (see Table \ref{TABLE-1}).
Increasing the number of splines improves slightly the precision.
However, extending for $N_{spline}=2$ the reciprocal summation using
$K_c=4 \pi/L$, the precision decreases to $0.1\%$, with a maximum error of $ 0.4\%$
(see Table \ref{TABLE-2}).
The short-range part of the optimized Coulomb potential writes
\begin{eqnarray}
  w(r) =\frac{q}{r}
 \left\{
 \begin{array} {l@{\quad : \quad}l}
  \sum_{i=0}^6 a_i^< \left(\frac{4 r}{L} \right)^i  &  0 \leq r < L/4 \\
 \sum_{i=0}^6 a_i^> \left(\frac{4 r}{L}-1 \right)^i  &  L/4 \leq r < L/2 
\end{array}
\right.
\label{SR-fast}
\end{eqnarray}
whereas the long-range part is  given by
\begin{eqnarray}
  y(r)  =  \frac{q}{L}\left\{ \tilde{y}_{0}
  + 2  \tilde{y}_{1} \left[ \cos( \hat x)+\cos( \hat y) \right]
  +4\tilde{y}_{2} \cos( \hat x)\cos( \hat y) 
  +2\tilde{y}_{4} \left[ \cos( 2\hat x)+\cos( 2\hat y) \right]
  \right\}
\label{LR-fast}
\end{eqnarray}
where $\hat x=2 \pi x/L$, $\hat y=2 \pi y/L$. See Fig.\ref{FIG-3} for a comparison of these simple expression with the ``exact'' periodic potential.

Both explicit expressions are roughly one order of magnitude better than the
corresponding Ewald potentials. Even if not extremely precise, the expressions should
extrapolate much better to the thermodynamic limit than any truncated potential using
only nearest image convention in real space. Further the real space part of the
optimized potential vanishes at $R_c=L/2$ by construction without introducing any
discontinuity in the potential and the derivatives at this point.

\begin{table}
\begin{center}
\begin{tabular}{|c||c|c|c|c|c|c|}\hline
 $i$ & 0 & 1 & 2 & 3 & 4 & 5 \\\hline
 $a_i^<$ & 1 & -0.819506 & 0 & 0.169304 & -0.146967 & 0.0777952  \\\hline
 $a_i^>$ & 0.280626 & -0.510485 & 0.404063  & -0.955541 & 1.3377 & -0.556365 \\\hline
\hline
\hline
 $n^2$ & 0 & 1 &  &  && \\\hline
 $\tilde{y}_{n^2}$ & -1.11863 & 0.124098 &  & & &   \\\hline
\end{tabular}
\end{center}
\label{TABLE-1}
\caption{
Optimized Potential parameters, Eq.(\ref{SR-fast} and Eq.(\ref{LR-fast}).
Top : short range real space parameters for $N_{spline}=2$. 
Bottom :  reciprocal space parameters where  $n=kL/2\pi$.
The mean precision is about $2\%$. 
}
\end{table}

\begin{table}
\begin{center}
\begin{tabular}{|c||c|c|c|c|c|c|}\hline
 $i$ & 0 & 1 & 2 & 3 & 4 & 5 \\\hline
 $a_i^<$ & 1 & -1.09583 & 0 & 0.30778 & -0.0359887 & -0.0266302  \\\hline
 $a_i^>$ & 0.149336 & -0.449592 & 0.441105 & -0.119121 & -0.0333851 & 0.0116568 \\\hline
\hline
\hline
 $n^2$ & 0 & 1 &  2 &  4  &&\\\hline
 $\tilde{y}_{n^2}$ & -0.870938 & 0.262177 & 0.0715766& 0.00474028 && \\\hline
\end{tabular}
\end{center}
\label{TABLE-2}
\caption{
Same as Table \ref{TABLE-1}.
The mean precision of this model is about $0.1\%$. 
}
\end{table}

\section{Conclusion}
We have shown that for the two dimensional Coulomb potential,
the numerically optimized potential can obtain a much higher precision compared
to the analytical Ewald potential summing over the same number of terms 
in reciprocal space. Therefore, the computational effort for many-body simulations involving
long range potentials can be significantly reduced using an optimized potential. 

For a pair potential the computational cost to evaluate the real space contribution to
the total potential is $\sim N N_{c}/2$ where $N$ is the total number of particle, $N_{c}=\pi R_c^2 \rho $ is the ``number of close neighbors''  
and $\rho=N/V$ the mean particle density. Since the number of $k-$vectors increases
as the volume in reciprocal space, the cost of the Fourier summation is roughly
$\sim N \pi K_c^2 V/4\pi^2$, and there is an optimum value for each system size, which
scales as 
$R_c \sim K_c^{-1} \sim N^{1/4}\sim L^{1/2}$
in the limit 
of a large particle number,
so that the computational cost in reciprocal space $\sim N^{3/2}$ roughly equals that
in real space.
We further note, that  in the limit of a large system, 
the total cost for evaluating the Coulomb potential using real and reciprocal space
summations is always favorable compared to any truncated potential with minimum image convention,
which scales as  
$\sim N^2$. 

Apart from a potential speed-up of simulations involving charged particles,
the big advantage of the optimized potential is its flexibility to split-up any
(long-ranged) function into a real-space and a reciprocal space contribution. 

\section*{Appendix A}
In this Appendix, we describe how to evaluate precisely the integral of 
the Bessel function $J_0(x)$.
Beginning with the evaluation of the Bessel function $J_{0}$, three domains are defined.
Around the origin, $0 \leq x \leq x_1$ the Bessel function is accurately evaluated
form the absolutely convergent series representation 
\cite{Abramowitz}
\begin{equation}
J_0(x)=\sum_{k=0}^\infty  \left( - \frac{x}{2 k!} \right)^{2k}, \quad 0 \leq x \leq x_1
\end{equation}
With standard double precision, a precision of  $10^{-14}$ is obtained up to $x_{1}=10$ by summing all terms whose absolute value is larger than  $10^{-18}$.
For large arguments, $x_2 \leq x < \infty$, the asymptotic expansion is used:
\cite{Abramowitz}
\begin{equation}
J_0(x)=\sqrt{\frac{2}{\pi x}} \left[ P(x) \cos(x-\pi/4) + Q(x) \sin(x-\pi/4) \right],
\quad x_2 \leq x < \infty
\end{equation}
with
\begin{eqnarray}
P(x) =  1+ \sum_{k=1} \frac{(-1)^k}{(8x)^{2k}} 
\prod_{m=0}^{2k-1} \frac{(2m+1)^2}{(m+1)}, &&
Q(x) =   \sum_{k=0} \frac{(-1)^k}{(8x)^{2k+1}} 
\prod_{m=0}^{2k} \frac{(2m+1)^2}{(m+1)} 
\end{eqnarray}
As usual, the summation is stopped when the absolute value of the running term of the series starts to increase. With the asked precision of $10^{-14}$, one finds $x_{2}=16$.
In the intermediate region $x_1<x<x_2$, the Chebyshev-pade approximant are calculated using Maple.
This strategy can be extended to any desired precision by cutting this interval in pieces.

The series expressions for $J_0(x)$ are then analytically integrated to
calculate $\int dx \, J_0(x)$. Unfortunately, the asymptotic expression is less converging, giving $x_{2}=30$ for a precision of $10^{-14}$. The Chebyshev-pade approximant in the interval $[10,30]$ is calculated at order 36, thanks to Maple:
{\sl
with(orthopoly):
with(numapprox):
Digits:=20;
IJ:=int(BesselJ(0,x),x);
IJCh:=eval(chebpade(IJ, x=10..30,36)):
convert(subs(x=10*(X+2),IJCh),horner);
}
where $X=(x-20)/10$. The {\sl gsl} routine have been used for $J_{0}$ and $J_{1}$\cite{GSL}.

\section*{Appendix B}
It is also possible to fix the constant term at the origin in the optimized Coulomb potential,
$t_{01}$ in Eq.(\ref{wr}), by imposing the Madelung constant of the underlying lattice,
$v_{Mad}=lim_{r \to 0} \left( v_{PP}(r) - q/r \right)$.
Using the Ewald expressions, Eq.(\ref{ewald}), the Madelung constant writes
\begin{equation}
v_{Mad}=\sum_{\lv \ne (0,0)} w^\alpha(\lv) + \sum_{\kv} \tilde{y}_\kv^\alpha
-2 \sqrt{\frac{\alpha}{\pi}}
\end{equation}
which can be calculated with high precision choosing
$\alpha=\sqrt{\pi}/L$;
for the square lattice $v_{Mad}=-3.90026492000195 q/L$.
For the optimized
potential, imposing the $1/r$ divergency with $C=1$,
the Madelung verifies $v_{Mad}=t_{01}+\sum_{k\leq K_c} \tilde{y}_k$. Since $\tilde{y}_k$ is coupled
to all $t_{i\alpha}$ by Eq.(\ref{yk}), this constraint leads to modifications
of Eq.(\ref{LE_constr}),
\begin{equation}
\sum_{(j,\beta) \ne \{(0,0),(0,1),(0,2)\}} A_{i\alpha,j\beta}  t_{j \beta} = b_{i\alpha},
\quad 0\leq \alpha \leq m, \; 0\leq i \leq N_{spline}
\end{equation}
for
$(i,\alpha) \ne \{(0,0),(0,1),(0,2)\}$
with
\begin{eqnarray}
A_{i\alpha,j\beta}&=&\sum_{k=K_c}^{K_m} 
 \tilde{\tilde{c}}_{i\alpha k}
 \tilde{\tilde{c}}_{i\beta k}
\\
b_{i \alpha}&=&\sum_{k=K_c}^{K_m} 
\left( \tilde{\tilde{v}}_{k}
- \tilde{c}_{00k} t_{00} - \tilde{c}_{10k} \frac{v_{Mad}}{1-\sum_{q\leq K_c} \tilde{c}_{10q}} 
-\sum_{j \beta\ne (0,1)} t_{j\beta} (\tilde{\tilde{c}}_{j\beta q} -\tilde{c}_{j\beta q} )
\right) 
\tilde{\tilde{c}}_{i \alpha k}
\label{LE_constr2}\\
  \tilde{\tilde{X}}_{k}&=& \tilde{X}_{ k}
+\tilde{c}_{10 k} \frac{\sum_{q\leq K_c} \tilde{X}_{q} }{1-\sum_{q\leq K_c} \tilde{c}_{10q}}
\end{eqnarray}
instead of Eq.(\ref{LE}).
Here $X$ denotes either $c_{i\alpha}$ or $v$.
Including the Madelung term as
a constraint improves slightly the solution.

ACKNOWLEDGMENTS: We acknowledge discussions with D. Ceperley.
M.H. is grateful to S. Chiesa for providing the 3D-version of the optimized potential
used for comparisons.

 \end{document}